# Preservación de la memoria colectiva-científica, en la astronomía argentina, desde el Observatorio de La Plata

Natalia Soledad Meilán[1] , Yael Aidelman[1,2], Lydia Cidale[1,2], Roberto Gamen[1,2], Mónica López[1], Romina Peralta Pascula[1]

[1] Facultad de Ciencias Astronómicas y Geofísicas (FCAG), Universidad Nacional de La Plata (UNLP), Argentina.
[2] Instituto de Astrofísica de La Plata (CONICET-UNLP)

**Resumen:** Este trabajo describe el rescate y puesta en valor de varios de miles de placas espectrográficas del Observatorio de La Plata. Estas placas y sus documentos adjuntos, escritos de puño y letra, datan de principios del siglo XX, y contienen información de observaciones realizadas en diferentes observatorios del mundo. Debido a su fragilidad y al inevitable deterioro por el paso del tiempo, están siendo saneadas, recuperadas y digitalizadas. La información contenida en las placas tiene un alto valor científico. Asimismo la comparación de un espectro antiguo con uno actual, permite reconstruir parte de la historia de la estrella. A través del proyecto ReTrOH no sólo se busca traducir el documento a un formato adecuado, preservar y difundir esos conocimientos, sino también recuperar el proyecto científico y el contexto histórico y personal de los observadores. Los espectros digitalizados se encuentran disponibles en el repositorio virtual (SEDICI) de la UNLP.

**Abstract:** This work describes the rescue and enhancement of several thousand spectrophotometric plates from the La Plata Observatory. These plates and their attached handwritten documents, date from the beginning of the 20th century and contain information on observations made at different observatories around the world. Due to their fragility and the inevitable deterioration produced by the passage of time, they are being cleaned up, recovered and digitized. The information contained in the plates has a high scientific value and, based on a comparison of an old spectrum with a current one, part of the history of the star can be reconstructed. Through the ReTrOH project, we seek to translate the document into an appropriate format, preserve and disseminate that knowledge, and also recover the scientific project and the historical and personal context of the observers. The digitized spectra are available in the virtual repository (SEDICI) of the UNLP.

## 1. Introducción:

El Observatorio de La Plata, incorporado actualmente a la Facultad de Ciencias Astronómicas y Geofísicas (FCAG), de la Universidad Nacional de La Plata (UNLP), cuenta con un importante acervo patrimonial de placas fotográficas de eventos históricos y de diferentes objetos del cielo austral, como estrellas, planetas y cometas, que datan desde principios del siglo XX.
Las placas fotográficas cuentan una parte de la historia de la astronomía argentina. Ellas son testigos de las distintas campañas y misiones internacionales en las que participó nuestro país: la observación del tránsito de Venus (1882); las campañas de seguimiento de eclipses total de sol en pos de confirmar la Teoría General de la Relatividad (1914-1919), y de la relevancia de los proyectos y trabajos científicos desarrollados en la República Argentina (Paolantonio, 2021a; Paolantonio, 2021b; Paolantonio et al., 2019; Rieznik, 2013; Bernaola, 2004, entre otros).

El proyecto ReTrOH[1] (REcuperación del TRabajo Observacional Histórico), nace en 2019, poco después de la Creación del Repositorio Científico de la FCAG (Res. Nº 169/19) con el objetivo de rescatar, conservar, recuperar, digitalizar y poner en valor tanto las placas fotográficas como así también los demás datos científicos adquiridos en el Observatorio desde su fundación (el 22 de noviembre de 1883). Su origen fue motivado por la Resolución Nro. B3 *"Safeguarding the information in photographic plates"* emitida por la Unión Astronómica Internacional (IAU) del año 2000, donde se solicita a la comunidad astronómica que se tomen medidas para conservar los datos históricos y se realice la transferencia de datos analógicos a formatos digitales, y por la creación del Nuevo Observatorio Virtual Argentino (NOVA), en 2009.

Desde el año 2000, en distintas partes del mundo se ha comenzado con la labor de digitalización de placas fotográficas, por ejemplo: la colección de placas astronómicas de Harvard[2], del Observatorio Maria Mitchell[3], o la colección de Placas Astronómicas Digitalizadas de Heidelberg[4], por mencionar algunas. Particularmente en Argentina, por esos años, se comenzaron a digitalizar placas fotográficas creando el

1 https://retroh.fcaglp.unlp.edu.ar
2 **http://dasch.rc.fas.harvard.edu/project.php**
3 **https://www.mariamitchell.org/astronomical-plates-collection**
4 **https://dc.zah.uni-heidelberg.de/hdap**

Primer archivo digital de placas fotográficas del Observatorio Astronómico de Córdoba (Calderón et al., 2004) en el cual se incluyen las placas pertenecientes al proyecto *Carte du Ciel*, entre otras.

## 2. Las Placas de Vidrio

Las placas de vidrio se comenzaron a utilizar a mediados del siglo XIX, manteniéndose vigentes hasta mediados del siglo XX. Las mismas se utilizaron como soporte para la fotografía. Si bien, a fines del siglo XIX comenzaron a surgir nuevos soportes, menos frágiles, pesados y voluminosos, el trabajo profesional en el ámbito de la astronomía continuó desarrollándose en placas de vidrio hasta la década de los ’90 (siglo XX), ya que éstas presentaban gran estabilidad y un alto grado de resolución, ofreciendo una calidad superior, comparadas con el soporte en película. La fotografía convencional fue finalmente desplazada y reemplazada rápidamente por la fotografía digital ante el surgimiento de los detectores CCD (dispositivos de carga acoplada).

La colección de placas espectrográficas del Observatorio de La Plata data de 1904. Estas fueron adquiridas en distintos observatorios del mundo, entre ellos podemos mencionar el Observatorio de La Plata (actual FCAG), la Estación Astrofísica de Bosque Alegre (Córdoba), el Observatorio Interamericano de Cerro Tololo (CTIO, Chile), el Observatorio de Mount Wilson (USA), entre otros. Sus tamaños son muy diversos: 4 x 6 cm, 7 x 2 cm, 8.80 x 0.4 cm, 10 x 2.6 cm, etc. El tamaño, en general, está definido por las dimensiones del porta placa o chasis del espectrógrafo, pero como las mismas eran cortadas por los observadores, sus medidas no son uniformes. En algunos casos, las placas están colocadas en un soporte de plástico o fueron pegadas sobre otro soporte de vidrio de mayor tamaño para facilitar su manipulación. Cada placa se encuentra protegida por un sobre de papel o cartón, como se muestra en la Figura 1.

Uno de los objetivos del proyecto ReTrOH es realizar una conservación preventiva del material. Gaël de Guichen (1995, 1999) define a este proceso como "el conjunto de las acciones destinadas a asegurar la salvaguarda (o a aumentar la esperanza de vida) de una colección o de un objeto". En este contexto, se describen a continuación las tareas de rescate, saneamiento, digitalización y puesta en valor de la colección de placas que se hallan depositadas en el recinto que aloja al Microdensitómetro Grant, ubicado en el sótano de la Facultad. Además de realizar una conservación preventiva del material, se busca reunir información sobre los proyectos científicos, grupos de investigadores y observadores; recuperar la información científica contenida en las mismas; profundizar en la importancia de digitalizar registros de espectros en placas de vidrio y ofrecer el material a la comunidad a través de los repositorios del sistema nacional. Además, estos objetivos se ven motivados ante el hecho de que las mediciones realizadas con métodos modernos sobre las imágenes digitalizadas resultan más precisas que las utilizadas antiguamente (Davis et al., 2004).

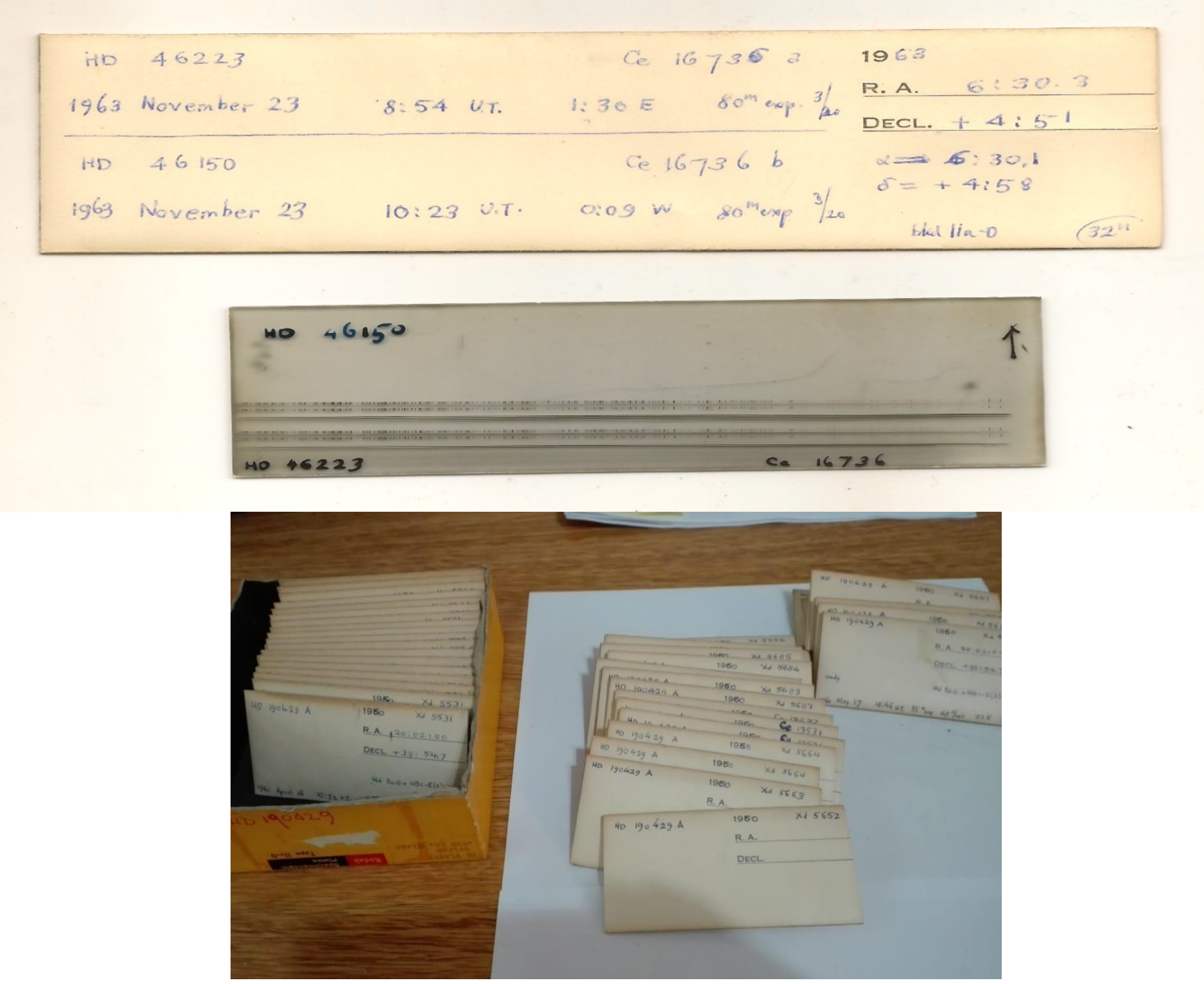

**Figura 1.** Arriba: sobre y placa de vidrio con el espectro de dos estrellas HD 46150 y HD 46223. También se observa el espectro de la lámpara de comparación. Abajo: caja y sobres que contienen a las placas de vidrio.

La FCAG tiene en formación un archivo histórico de espectros estelares registrados en placas de vidrio. Relevar, investigar, digitalizar, y así poner en valor, las placas de vidrio de investigaciones realizadas durante décadas por astrónomos y astrónomas de nuestro país, no solo tiene el valor intrínseco de documentar muchos descubrimientos realizados en La Plata, sino que tiene un valor agregado para futuros investigadores e historiadores de la ciencia, pues los espectros aportan información sobre: temperatura, gravedad, abundancias, componente radial de la velocidad, rotación, etc. Los mismos se pueden considerar como el ADN de las personas, dejan rastros de la historia de sus objetos, preservan información importante o “descubrimientos latentes” que aún no han sido revelados.

Es interesante aquí enumerar importantes trabajos científicos que han surgido de estudiar y reanalizar la información contenida en placas de vidrio. Por ejemplo, utilizando datos de placas digitalizadas, almacenadas en la Universidad de Hamburger Sternwarte, Alemania, se detectaron variaciones de largo período en blazares (Wertz et al., 2017); en el Instituto Astronómico de la Academia de Ciencias de Uzbekistán se obtuvieron posiciones y magnitudes estelares (Muminov et al., 2017); en el Observatorio de Sonneberg, Alemania, se construyeron curvas de luz de estrellas T Tauri (Heines, 1999); con datos del Digital Access to a Sky Century @ Harvard (DASCH1), Walborn et al. (2017) descubrieron eventos eruptivos en la estrella Variable Luminosa Azul R71 ocurridos a comienzos del siglo pasado y nunca reportados.

En nuestro país, utilizando datos astrométricos derivados a partir de cuatro placas *Carte du Ciel* y una del Catálogo Astrográfico de la colección del Observatorio Astronómico de Córdoba, Orellana et al. (2010) detectaron un cúmulo abierto entre las estrellas de campo de la región de Collinder 132 y calcularon el movimiento propio medio y las probabilidades de pertenencia de las estrellas de la región.

Por otra parte, la información contenida en las placas tiene un alto valor científico ya que de la comparación de un espectro antiguo con uno actual, se puede reconstruir parte de la historia de la estrella.

Las placas de vidrio de espectros estelares del Observatorio de La Plata están siendo puestas en valor mediante un análisis interdisciplinar, por parte de astrónomas, astrónomos y museólogas, para así obtener información resultante de observaciones realizadas durante casi un siglo. Contextualizarlas históricamente, ponerlas en valor de forma integral, i.e. tanto la conservación física de las placas de vidrio como el conocimiento que en ellas se deposita, son la razón fundamental para su rescate, mediante su digitalización. El objetivo es que el acceso a esa información sea universal y que contribuya a la conservación preventiva del material original que da cuenta de algunas de las prácticas científicas de este Observatorio.

## 3. Metodología

Entendiendo la relevancia que conlleva recuperar la información que contienen estas placas de vidrio, y sus correlatos en documentos escritos, se procedió a realizar un estudio ambiental del depósito que las aloja. Se llevó a cabo, además, el rescate de un grupo de placas y documentos en papel asociados a las mismas, iniciando así su puesta en valor. Asimismo se comenzó con las tareas de preservación preventiva, digitalización e inventario de cada placa, poniendo así dicho repositorio a disposición de la comunidad educativa y científica.

### 3.1. Estudio Ambiental y saneamiento

Los espacios físico-arquitectónicos que actualmente ocupa la colección de los datos espectroscópicos están ubicados en el subsuelo del edificio central de la FCAG, ala sudeste, en un recinto de 16 $m^2$, donde se encuentra alojado el microdensitómetro Grant (ver Figura 2).
En 2019, se procedió a realizar estudios ambientales del recinto y del estado del material depositado allí. El mismo estuvo a cargo de investigadores y docentes del Instituto Spegazzini (UNLP) y del Instituto Biológico-Entomológico de la Facultad de Ciencias Biológicas y Museo de La Plata, quienes emitieron un informe micológico y entomológico, brindando además recomendaciones de saneamiento. Participaron también profesionales del Centro de Estudios Parasitológicos y de Vectores (CIPAVE, UNLP-CONICET), emitiendo un diagnóstico de artrópodos plaga (el cuál reveló la presencia de una gran cantidad de arañas de tipo *loxosceles laeta*, altamente venenosas).
Estos estudios se llevaron a cabo con la finalidad de realizar un saneamiento ambiental adecuado para un rescate óptimo de las placas y documentos asociados, así como para la seguridad de los profesionales que realizan dicha tarea.
Se prevé un plan de saneamiento del recinto del sótano, actividad que fue pospuesta por la situación de pandemia durante los años 2020 y 2021.

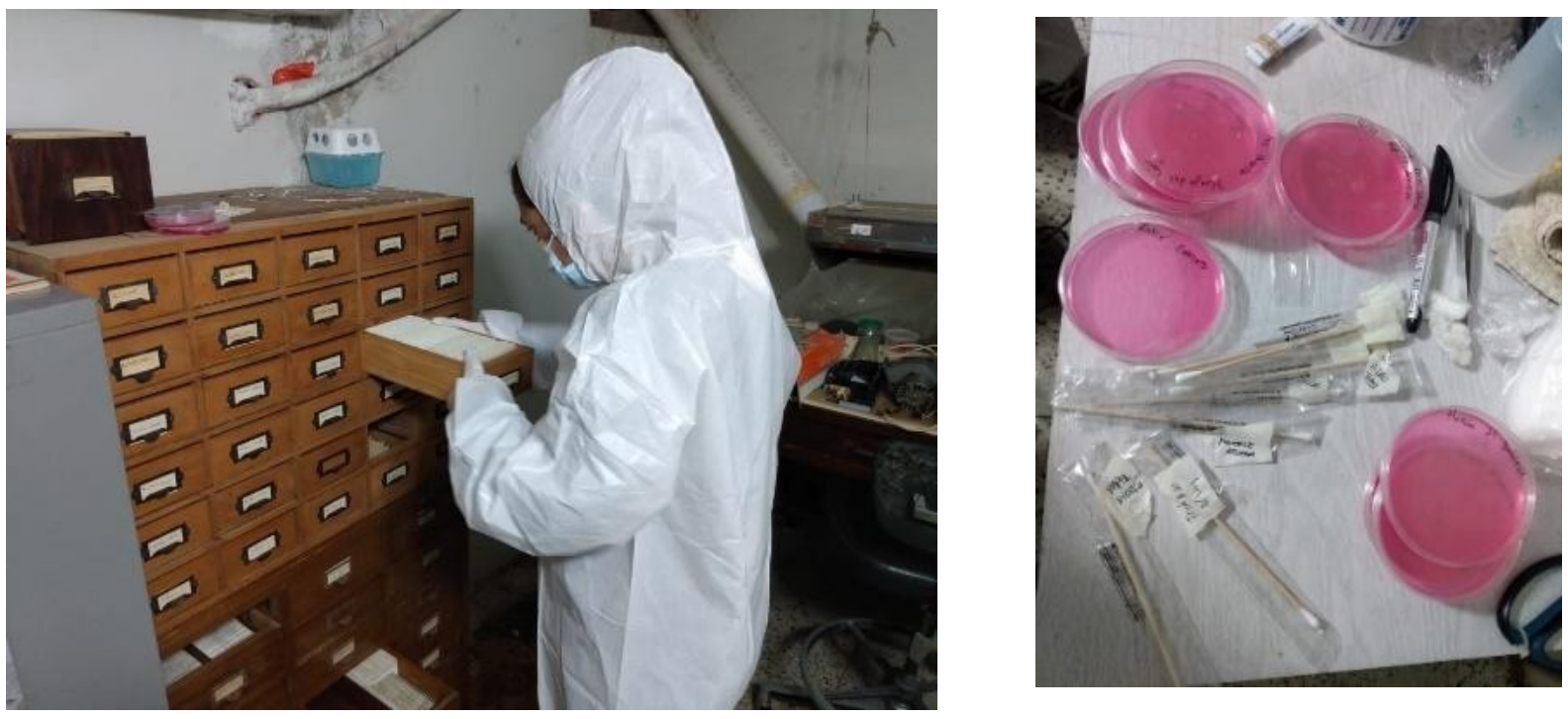

**Figura 2:** Estudios ambientales del recinto Grant y de la muestra.

### 3.2. Rescate y conservación preventiva

Las placas a recuperar son trasladadas a un nuevo espacio físico, un espacio apartado del ambiente general de trabajo: el taller de conservación o la oficina, y son puestas en cuarentena. Este nuevo ambiente es algo más propicio que el del subsuelo, donde los niveles de humedad y las colonias de hongos son altas.

Quienes desempeñan esa tarea usan indumentaria adecuada: enterito protector y/o delantal, antiparras, barbijo y guantes de látex. La finalidad de todo este proceso es evitar contaminar y contaminarnos, haciendo de la asepsia un principio básico de la conservación preventiva.

Estas placas se encuentran en cajas con denominaciones hechas de puño y letra de las astrónomas y astrónomos que llevaron a cabo las observaciones.

El equipo interdisciplinario del grupo ReTrOH ha consensuado en respetar el orden en que el material fue hallado, conservar la nomenclatura y la secuencia en que se encuentran en dichas cajas. Por lo tanto, esta es la forma en que se organizará el archivo, respetando su origen.

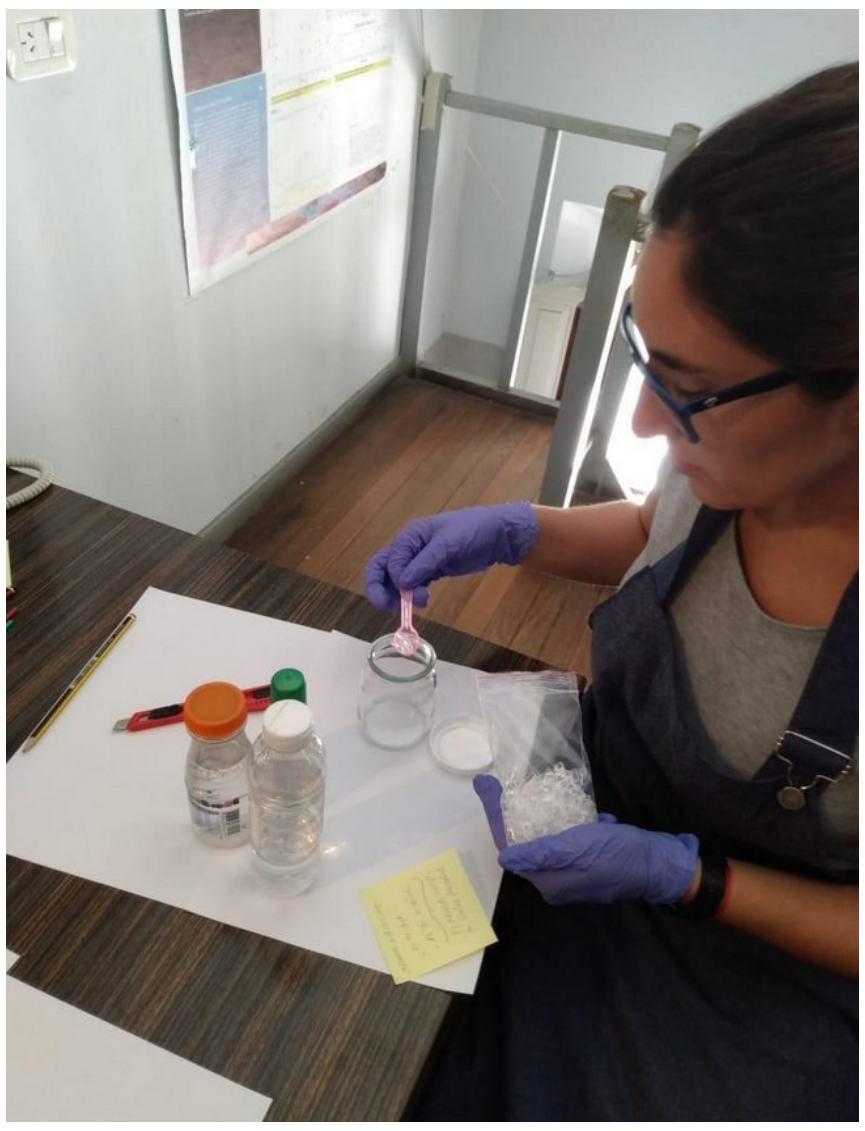

**Figura 3:** Tareas de limpieza y saneamiento.

Los pasos realizados para el procedimiento de rescate y preservación preventiva fueron:

1) Retirar las placas de la caja original (se fue haciendo caja por caja);
2) Realizar una limpieza mecánica y manual. Las placas fueron extraídas una por una de su sobre contenedor y se procedió a la limpieza del sobre con pinceleta suave, dejándolo libre del polvo. Las huellas de hongos causadas por la humedad ambiental sufrida durante todos estos años no han podido removerse por completo, pero se hallan inactivadas luego de un período de cuarentena. El cambio de ambiente y la estabilización gradual de la humedad ocurridos durante la fase de cuarentena, evita así un resecamiento brusco que pudiera resquebrajar los sobres que son material histórico, como así también las placas, ya que son un documento que da cuenta de cómo se trabajaba en otras épocas;
3) Las placas también son pinceladas suavemente sobre un campo de cartulina libre de ácido o tela de algodón neutro, sólo del lado opuesto a la emulsión, dado que en ella está grabada la imagen y, en la instancia actual del proyecto, aún no fue intervenida. Sólo se estabiliza, como en el caso de los sobres, durante la cuarentena, evitando la proliferación de hongos, que ya no accionan al haber pasado por tramos de deshumificación mecánica;
4) Una vez concretada la limpieza de ambas piezas, la placa vuelve al sobre y ambos son guardados en la caja original, que también se limpia de tierra, desechos varios como ganchos metálicos, restos de banditas elásticas resecas que son un elemento con compuesto orgánico con un alto grado de deterioro y que compromete el estado de las piezas que se encuentran asociadas.

Una vez llevadas las placas fuera de espacios de subsuelo, los arácnidos no proliferan y son controlados con desinfecciones periódicas.

### 3.3. Digitalización

Otra etapa importante para la conservación preventiva y recuperación de la información impresa en la emulsión de las placas fotográficas, sin poner en peligro su integridad, es la digitalización, tanto de las placas propiamente dichas como así también de sus sobres contenedores y los documentos adjuntos correspondientes (como cuadernos con anotaciones y planillas de observación). De este modo es posible la conservación científica-astronómica, preservando los datos precisos que estas placas tienen, como así también la museológica-histórica, dado que los sobres, las placas y los documentos adjuntos dan información sobre la forma de archivar y racionalizar los contenedores de datos científicos de la época y dan cuenta de la contextualización témporo-histórica de esta colección.

La posibilidad de disponer de los datos históricos en formato digital permite, no sólo comparar dichas observaciones con las actuales, para detectar variaciones de intensidad y en velocidad radial de las diversas fuentes astronómicas observadas durante varias décadas, sino también para analizar dichas observaciones con herramientas y códigos modernos. Para este fin, es necesario, entonces, digitalizar, procesar y convertir las placas a un formato estándar actual, como por ejemplo el formato FITS.

El primer paso del proceso de digitalización consiste en el escaneo de las placas espectrográficas. De este modo, se transforman las placas analógicas en archivos de imagen en formato TIFF. Esta tarea se está llevando a cabo utilizando un escáner Nikon 9000ED (optimizado para placas fotográficas espectrales), proporcionado por el Observatorio Virtual Argentino (NOVA, CONICET), empleando una resolución de 4000 dpi. Luego, utilizando un software escrito en lenguaje PYTHON[5], estos datos son transformados al formato FITS. El código también permite incorporar al *"header"* (encabezado de la imagen) información referente al objeto observado, al instrumental y configuración utilizados, y al Observatorio desde el cuál se adquirió el espectro. Paralelamente, empleando un escáner de documentos se procede a la digitalización del sobre que contiene a la placa.

Esta tarea comenzó en el año 2019 pero se vió interrumpida por la pandemia, así que en la actualidad sólo contamos con unas 200 placas digitalizadas. Los archivos FITS se suben a un repositorio virtual (SEDICI[6], UNLP) otorgándole un acceso universal que completa la puesta en valor y la conservación preventiva de estos documentos, ubicándolos en el campo del acceso público, siendo estos documentos históricos pertenecientes al ámbito de lo público por origen. La consulta digital posibilita esta conservación preventiva ampliando horizontes y evitando la manipulación de tan valiosas y frágiles piezas científicas.

Durante esta etapa se realizó un diagnóstico preliminar del estado de conservación de cada una de las placas para poder hacer el análisis cuanti-cualitativo que se verá más adelante. También se fue realizando el Inventario Somero.

Actualmente, también se están realizando las tareas de prevención y saneamiento de los cuadernos de observación. Una vez recuperados y digitalizados los mismos serán incorporados al SEDICI.

5 Este software fue desarrollado por integrantes del grupo ReTrOH y mejorado recientemente por colaboradores de la Facultad de Informática de la UNLP.

6 http://sedici.unlp.edu.ar/handle/10915/74499/discover?sort_by=dc.date.accessioned_dt&order=desc

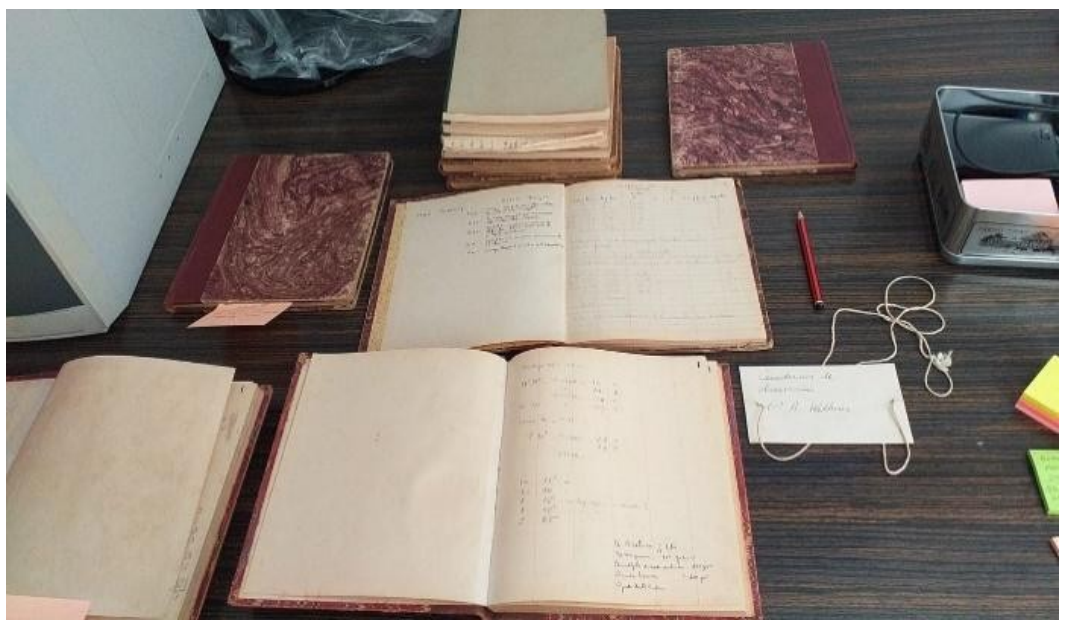

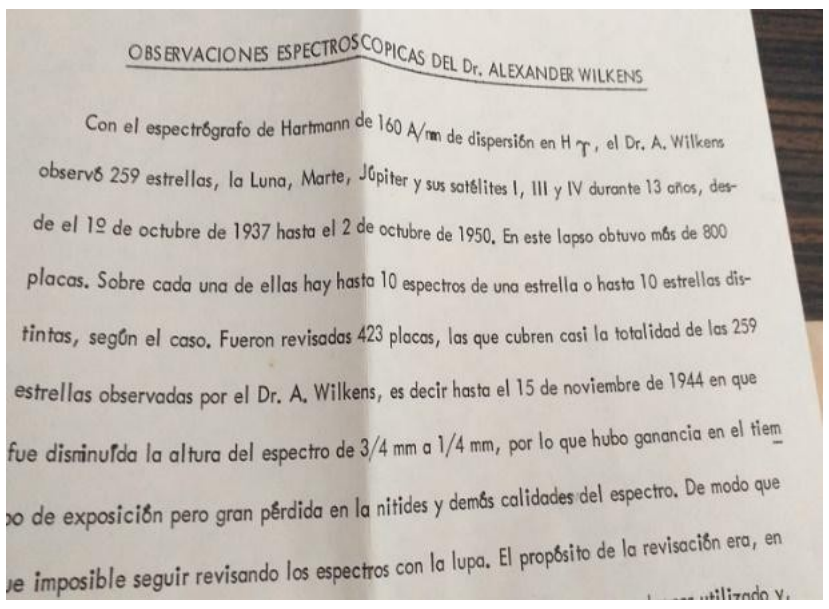

OBSERVACIONES ESPECTROSCOPICAS DEL Dr. ALEXANDER WILKENS

Con el espectrógrafo de Hartmann de 160 A/mm de dispersión en H γ, el Dr. A. Wilkens observó 259 estrellas, la Luna, Marte, Júpiter y sus satélites I, III y IV durante 13 años, desde el 1º de octubre de 1937 hasta el 2 de octubre de 1950. En este lapso obtuvo más de 800 placas. Sobre cada una de ellas hay hasta 10 espectros de una estrella o hasta 10 estrellas distintas, según el caso. Fueron revisadas 423 placas, las que cubren casi la totalidad de las 259 estrellas observadas por el Dr. A. Wilkens, es decir hasta el 15 de noviembre de 1944 en que fue disminuída la altura del espectro de 3/4 mm a 1/4 mm, por lo que hubo ganancia en el tiem-
po de exposición pero gran pérdida en la nitides y demás calidades del espectro. De modo que
ue imposible seguir revisando los espectros con la lupa. El propósito de la revisación era, en

**Figura 4.** Observaciones espectroscópicas realizadas por el Dr. Alexander Wilkens e informe correspondiente emitido por el Sr. Boris Kucewicz en 1964.

## 4. Verificación de la calidad de los datos

La verificación de la calidad de los  espectros digitalizados se realizó empleando las observaciones de la estrella HD 50845 (Meilán, 2018). Los espectros originales en soporte de vidrio fueron adquiridos en 1984 por los Dres. Jorge Sahade y Adela Ringuelet en el CTIO (Chile) utilizando un espectrógrafo Coudé y publicados por Sahade & Ringuelet (1985) y Sahade et al. (1987). Meilán (2018), en su trabajo de tesis de licenciatura, desarrolló un método adecuado para digitalizar estos espectros transformándolos a archivos en formato FITS. Luego, realizó la extracción y la calibración en longitud de onda. Finalmente, analizó los espectros digitalizados empleando herramientas actuales como las tareas del código IRAF (*Image Reduction and Analysis Facility*, desarrollado por *the National Optical Astronomy Observatory, NOAO*). Los resultados obtenidos referentes a velocidad radial medida y el tipo espectral de la estrella fueron comparados con los publicados en el trabajo original. El excelente acuerdo encontrado permitió validar el procedimiento de digitalización y reducción empleados (para más detalles ver Meilán, 2018, y Meilán et al., 2020).

## 5. Conclusiones

Poner en valor esta importante colección de datos astronómicos y hacerla accesible de forma universal, es necesario para seguir construyendo esta parte de la historia y la memoria colectiva y científica de nuestro país, como así también de la Facultad de Ciencias Astronómicas y Geofísicas de la Universidad Nacional de La Plata.

Relevar, analizar, digitalizar y poner en valor las placas espectrográficas, en las cuales los investigadores han plasmado y documentado diferentes fenómenos celestes durante décadas, es nuestra responsabilidad. Heredamos un valioso patrimonio, el cual puede utilizarse en futuras investigaciones tanto científicas, ya que en las placas hay descubrimientos latentes, como para quienes se dedican a la historia de las ciencias.

A futuro, el proyecto ReTroH tiene un objetivo más amplio, como ser la recuperación de las placas astrométricas, y la recuperación de material histórico científico del área de geofísica y meteorología. Parte de esta tarea será en forma fotográfica, empleando máquinas réflex y una lente aplanadora de campo.

***Agradecimientos. Agradecemos al árbitro por sus sugerencias y comentarios. Este proyecto ha recibido financiación en el marco del Proyecto de Digitalización y Recuperación del Archivo de Placas Espectroscópicas de la Facultad de Ciencias Astronómicas y Geofísicas (PID 11/G167).***

**Bibliografía**